\documentclass[11pt,superscriptaddress,amssymb,amsmath]{revtex4}
\usepackage{txfonts}
\usepackage{color,latexsym,epsfig,bm,times,subfigure}
\usepackage[ps2pdf,bookmarks=true,colorlinks,linkcolor=blue,urlcolor=green,citecolor=red]{hyperref}

\usepackage{fancyhdr}
\renewcommand\b{\beta}
\renewcommand\d{\delta}

\renewcommand\l{\lambda}

\renewcommand\t{\tau}

\renewcommand\c{\chi}
\renewcommand\j{\psi}
\renewcommand\o{\omega}
\newcommand\e{\epsilon}
\newcommand\g{\gamma}

\newcommand\m{\mu}

\newcommand\x{\xi}
\newcommand\p{\pi}
\newcommand\h{\theta}
\newcommand\s{\sigma}

\newcommand\w{\eta}
\newcommand\ve{\varepsilon}

\renewcommand\S{\Sigma}
\renewcommand\O{\Omega}

\newcommand\D{\Delta}

\newcommand{\fig}[1]{Fig.~\ref{#1}}
\newcommand{\eq}[1]{Eq.~(\ref{#1})}
\newcommand{\eqs}[2]{Eqs.~(\ref{#1})-(\ref{#2})}

\newcommand\lb{\left(}
\newcommand\rb{\right)}
\newcommand\ls{\left[}
\newcommand\rs{\right]}

\newcommand{\lan}{\langle}
\newcommand{\ran}{\rangle}

\newcommand\ra{\rightarrow}

\newcommand{\non}{\nonumber\\}
\newcommand\pt{\partial}

\newcommand{\etal}{\emph{et al.}}

\newcommand{\ch}{{\cal H}}

\newcommand{\bx}{{\mathbf x}}

\newcommand{\bp}{{\mathbf p}}
\newcommand{\bk}{{\mathbf k}}

\newcommand{\bs}{{\bm{\s}}}
\newcommand{\bo}{{\bm{\o}}}

\newcommand{\bb}{{\bf B}}
\newcommand{\be}{{\bf E}}
\newcommand{\bL}{{\bf W}}
\newcommand{\ba}{{\bf A}}
\newcommand{\bl}{{\bf w}}

\newcommand{\bO}{{\bm \O}}
\newcommand{\bj}{{\bf j}}

\begin{document}

\title{Simulating Chiral Magnetic and Separation Effects with Spin-Orbit Coupled Atomic Gases}
\author{Xu-Guang Huang}
\affiliation{Physics Department and Center for Particle Physics and Field Theory, Fudan University, Shanghai 200433, China.}

\date{\today}

\begin{abstract}
{\bf The chiral magnetic and chiral separation effects---quantum-anomaly-induced electric current and chiral current along an external magnetic field in parity-odd quark-gluon plasma---have received intense studies in the community of heavy-ion collision physics. We show that analogous effects occur in rotating trapped Fermi gases with Weyl-Zeeman spin-orbit coupling where the rotation plays the role of an external magnetic field. These effects can induce a mass quadrupole in the atomic cloud along the rotation axis which may be tested in future experiments. Our results suggest that the spin-orbit coupled atomic gases are potential simulators of the chiral magnetic and separation effects.}
\end{abstract}

\maketitle

The recent experimental breakthroughs in generating synthetic spin-orbit coupling (SOC) in both bosonic~\cite{Lin:2011} and fermionic  gases~\cite{Wang:2012,Cheuk:2012} have opened a new era for cold atomic physics. In these experiments, a pair of Raman lasers induced an equal mixture of Rashba and Dresselhaus SOCs between the (psuedo)spin-$1/2$ internal degree of freedom and the orbital motion of the atoms. Very promisingly, other types of SOC, e.g., the Weyl SOC~\cite{Anderson:2012,Anderson:2013,Li:2012}, could also be realized. The presence of the SOC modifies the dynamics on both single-particle and many-body levels and a variety of novel properties have been explored. Furthermore, it is very suggestive that the spin-orbit coupled atomic gases may provide ideal platforms to simulate intriguing phenomena that have topological origins, e.g., the topological insulators or superfluid and Majorana fermions~\cite{Jiang:2011,Liu:2012,Zhang:2013,Ruhman:2015}, the spin Hall effect~\cite{Beeler:2013,Kennedy:2013}, and the Berezinskii-Kosterlitz-Thouless transition~\cite{He:2011,Xu:2015}. See Refs.~\cite{review:2013,review:2014,review:2015} for reviews.

In this article, we demonstrate that yet another topological phenomenon, the quantum anomaly, can also be realized in a special setup for the spin-orbit coupled atomic gases, namely the trapped rotating atomic gases with Weyl-Zeeman SOC. As a consequence of this quantum anomaly, the currents of opposite chiralities (see below for definition) are generated in parallel or anti-parallel to the rotation axis. These currents mimic the chiral magnetic effect (CME)~\cite{Kharzeev:2008,Fukushima:2008} and chiral separation effect (CSE)~\cite{Son:2004,Met:2005} that are intensively studied in the context of quark-gluon plasma (QGP) produced in heavy-ion collisions, with now the rotation playing the role of an external magnetic field.

The QGP version of the CME and CSE is expressed as
\begin{eqnarray}
\label{qcdcme}
\bj_V=\frac{e_fN_c\m_{A}}{2\p^2\hbar^2}\bb,\;\;\;\;
\bj_A=\frac{e_fN_c\m}{2\p^2\hbar^2}\bb,
\end{eqnarray}
for each flavor of light quarks, where $e_f$ is the electric charge of quark with flavor $f$, $\bj_V=\lan{\bar\j}{\bm\g}\j\ran$ and $\bj_A=\lan{\bar\j}{\bm\g}\g_5\j\ran$ are electric and chiral currents, $N_c=3$ is the color degeneracy, and $\mu$ and $\mu_A$ are vector and chiral chemical potentials. Experimentally, signals consistent with CME and CSE have been observed in heavy-ion collisions at Relativistic Heavy Ion Collider (RHIC)~\cite{STAR} and Large Hadron Collider (LHC)~\cite{ALICE}. In these collisions, extremely strong magnetic fields arise due to the fast motion of the ions~\cite{Skokov,Huang,Huang2015}, and these magnetic fields induce charge separation and chirality separation in the QGP via CME and CSE which in turn lead to special azimuthal distributions of the charged hadrons that are finally measured by the detectors.

It is worth noting that the CME was also discussed in astrophysical context~\cite{Vilenkin:1980} and more recently in Weyl and Dirac semimetals~\cite{Burkov:2012, Vaz:2013,Basar:2014,Land:2014,weyl:2014}. In Weyl and Dirac semimetals, the low energy excitations are Weyl and Dirac fermions. These materials can exhibit finite $\m_5$ due to their special band structure or by applying parallel electric and magnetic fields, and thus open the possibility of realizing the CME. Comparing to these previously explored systems, the atomic gases with Weyl-Zeeman SOC enable greater flexibility in controlling the parameters and thus provide not only simulators but also a unique mean to exploit new features (e.g., those raised by the presence of the Zeeman splitting field) of the CME and CSE.

{\bf Semiclassical approach.---}
We begin by considering the following single-particle Hamiltonian for spin-$1/2$ atoms (either bosons or fermions) in three dimensions (3D),
\begin{eqnarray}
\label{ham1}
\ch=\frac{\bp^2}{2m}-\m-\bL(\bp)\cdot\bs,
\end{eqnarray}
where $m$ is the mass, $\m$ is the chemical potential, and $\bp$ is the canonical momentum. The third term expresses a generic SOC where $\bs$ is the Pauli matrix and $\bL$ represents a momentum-dependent magnetic field. Its form will be given when necessary. The Hamiltonian (\ref{ham1}) represents a two-band system with band dispersions $\ve_c(\bp)=p^2/(2m)-\m-cW(\bp), c=\pm$. Correspondingly, we define the {\it chirality} of band $c$ to be right-handed (left-handed) if $c=+$ ($c=-$). We note that the chirality we defined here is commonly called {\it helicity} in literature. We choose the term ``chirality" to keep the consistence with the terminologies ``chiral anomaly", ``chiral magentic effect", etc.

Now consider that the atoms are trapped by an external potential $V(\bx)$ and at the meantime are subject to rotation with angular velocity $\bo$ which we assume to be a constant. The effect of the trapping and rotation can be described by a gauge potential $A^\m=(A_0,\ba)$ with $A_0(\bx)=V(\bx)-(m/2)(\bo\times\bx)^2-\m$ and $\ba(\bx)=m\bo\times\bx$, and then the Hamiltonian becomes
\begin{eqnarray}
\label{ham2}
\ch=\frac{[\bp-\ba(\bx)]^2}{2m}-\bL[\bp-\ba(\bx)]\cdot\bs+A_0(\bx).
\end{eqnarray}
Note that the Hamiltonian of atoms with SOC in the rotating frame is generally time dependent and the minimal substitution adoptted in going from \eq{ham1} to \eq{ham2} may not be applicable. However, for situations that, e.g., both the trap and the lasers inducing the SOC are rotating and the SOC is linear in momentum, one can find a time-dependent unitary transformation to eliminate the time dependence from the Hamiltonian; see Ref.~\cite{timedep} and the references therein. Nevertheless, the present study is restricted to such a situation and the use of the time-independent Halmitonian (\ref{ham2}) is justified.

We now derive a set of semiclassical equations of motion (EOM) for the orbital variables $\bx$ and $\bp$. (For non-rotating spin-orbit coupled atomic gases, a similar derivation is given in Ref.~\cite{bijl}.) At semiclassical level, we treat $\bx, \bp$, and $\bs$ as classical variables, and their EOM are easily derived from Hamiltonian (\ref{ham2}):
\begin{eqnarray}
\label{eomx1}
\dot{\bx}&=&\nabla_\bp\x-\nabla_\bp\bL\cdot\bs,\\
\label{eomp1}
\dot{\bp}&=&-\nabla_\bx\x+\nabla_\bx\bL\cdot\bs,\\
\label{eoms1}
\dot{\bs}&=&\frac{2}{\hbar}\bs\times\bL,
\end{eqnarray}
where $\x(\bx,\bp)=[\bp-\ba(\bx)]^2/(2m)+A_0(\bx)$ and $\bL(\bx,\bp)=\bL[\bp-\ba(\bx)]$. To proceed, we make an adiabatic approximation to the spin dynamics, that is, we treat the orbital degrees of freedom $\bx,\bp$ as slow variables while the spin $\bs$ as fast variable and solve \eq{eoms1} up to first order in time derivatives of the orbital variables. This gives
\begin{eqnarray}
\label{spin}
\bs\approx c\bl+c\frac{\hbar}{2W}\bl\times\dot{\bl},
\end{eqnarray}
where $\bl=\bL/W$. This procedure is essentially equivalent to solving $\bs$ up to first order in $\hbar$. Inserting \eq{spin} to \eqs{eomx1}{eomp1} we obtain
\begin{eqnarray}
\label{eomx2}
\dot{\bx}&=&\nabla_\bp(\x-cW)+c\hbar\bO_{px}\cdot\dot{\bx}+c\hbar\bO_{pp}\cdot\dot{\bp},\\
\label{eomp2}
\dot{\bp}&=&-\nabla_\bx(\x-cW)-c\hbar\bO_{xx}\cdot\dot{\bx}-c\hbar\bO_{xp}\cdot\dot{\bp},
\end{eqnarray}
where the tensors $\O_{AB}^{ij}$ with $A,B=x,p$ are defined as
\begin{eqnarray}
\O_{AB}^{ij}&=&\frac{1}{2}\lb\frac{\pt\bl}{\pt A_i}\times\frac{\pt\bl}{\pt B_j}\rb\cdot\bl.
\end{eqnarray}
In terms of the kinetic momentum,
$\bk\equiv\bp-\ba(\bx)$,
\eqs{eomx2}{eomp2} can be recast to more compact and transparently gauge invariant forms~\cite{Sundaram,Xiao:2010},
\begin{eqnarray}
\label{eomx5}
\dot{\bx}&=&\nabla_\bk\ve_c+c\hbar\dot{\bk}\times\bO,\\
\label{eomp5}
\dot{\bk}&=&\be+\dot{\bx}\times\bb,
\end{eqnarray}
where $\ve_c(\bk)=k^2/(2m)-\m-cW(\bk)$ and
\begin{eqnarray}
\label{elec}
E_i&=&-\frac{\pt A_0}{\pt x_i},\\
\label{magne}
B_i&=&\e_{ijk}\frac{\pt A_k}{\pt x_j}=2m\o_i,\\
\label{berrycur}
\O_i&=&\frac{1}{2}\e_{ijk}\O_{pp}^{jk},
\end{eqnarray}
are the effective electric field, magnetic field, and Berry curvature, respectively. Equation (\ref{magne}) exhibits the equivalence between $\bb$ and $\bo$. Note that for relativistic massless particles under rotation, the effective magnetic field would be momentum-dependent~\cite{Stephanov}, $\bb\sim 2|\bk|\bo$, and thus the rotation is no longer equivalent to a magnetic field. In that case, the rotation can induce an independent current other than the CME/CSE which is called chiral vortical effect (CVE)~\cite{chialve,chialve2,chialve3,Son2,cvw}. In the non-relativistic case, the CVE is equivalent to CME/CSE.

From \eq{eomx5} and \eq{eomp5} we obtain
\begin{eqnarray}
\label{eomx6}
\sqrt{G_c}\dot{\bx}&=&\nabla_\bk\ve_c+c\hbar\be\times\bO+c\hbar(\bO\cdot\nabla_\bk\ve_c)\bb,\\
\label{eomp6}
\sqrt{G_c}\dot{\bk}&=&\be+\nabla_\bk\ve_c\times\bb+c\hbar(\be\cdot\bb)\bO,
\end{eqnarray}
where $\sqrt{G_c}=1+c\hbar\bb\cdot\bO$, its physical meaning will be clear soon.
These are the semiclassical equations for the orbital motion of atoms of chirality $c$.
In these equations, the quantum effects are reflected in the Berry curvature terms.

We will hereafter focus on Fermi gases and we will use the natural units $\hbar=k_B=1$.

In the presence of the Berry curvature, the invariant measure of the phase space integration for atoms of chirality $c$ needs to be modified to $\sqrt{G_c} d^3\bx d^3\bk/(2\p)^3$~\cite{Xiao,dual:2006}. With this notification, one is able to write down a kinetic equation for the distribution function $f_c$ of chirality $c$,
\begin{eqnarray}
\label{kineq}
\pt_t f_c+\dot{\bx}\cdot\nabla_\bx f_c+\dot{\bk}\cdot\nabla_\bk f_c=I[f_c],
\end{eqnarray}
where $\dot{\bx}$ and $\dot{\bk}$ are given by \eqs{eomx6}{eomp6}. In the context of relativistic chiral medium like QGP, similar kinetic equation has been derived recently~\cite{Stephanov,Yamamoto,Wangqun,Wangqun2,Mannuel}. The density and current of chirality $c$ are defined as
\begin{eqnarray}
n_c&=&\int\frac{d^3\bk}{(2\p)^3}\sqrt{G_c}f_c,\\
\label{current1}
\bj_c&=&\int\frac{d^3\bk}{(2\p)^3}\sqrt{G_c}\dot{\bx}_cf_c,
\end{eqnarray}
and the continuity equation, following \eq{kineq}, reads
\begin{eqnarray}
\label{anomaly}
\pt_t n_c+\nabla_\bx\cdot\bj_c&=&c(\be\cdot\bb)\int\frac{d^3\bk}{(2\p)^3}f_c\nabla_\bk\cdot\bO \non
&=&cf_c(\bk_0)\frac{F}{4\p^2} \be\cdot\bb,
\end{eqnarray}
where we suppose that the collision kernel conserves the particle number for each chirality. If not, there would be a term $\int{d^3\bk}/{(2\p)^3}\sqrt{G_c}I[f_c]$ on the right-hand side. The $\bk_0$ in the second line specifies the location of the Berry monopole which coincides with the band-crossing point determined by $\bL(\bk_0)={\bf0}$. The $F$ is the total Berry curvature flux associated with the Berry monopole. Its explicit expression is
\begin{eqnarray}
F
&=&\frac{\e_{ijm}}{8\p}\int_{S^2}d^2\S_i
\bl\cdot\lb\frac{\pt\bl}{\pt k_j}\times\frac{\pt\bl}{\pt k_m}\rb.
\end{eqnarray}
This counts the winding number of the map $\bl: S^2\ra S^2$ and thus $F\in\p_2(S^2)=\mathbb{Z}$.

\begin{figure}
\begin{center}
\includegraphics[width=8.cm]{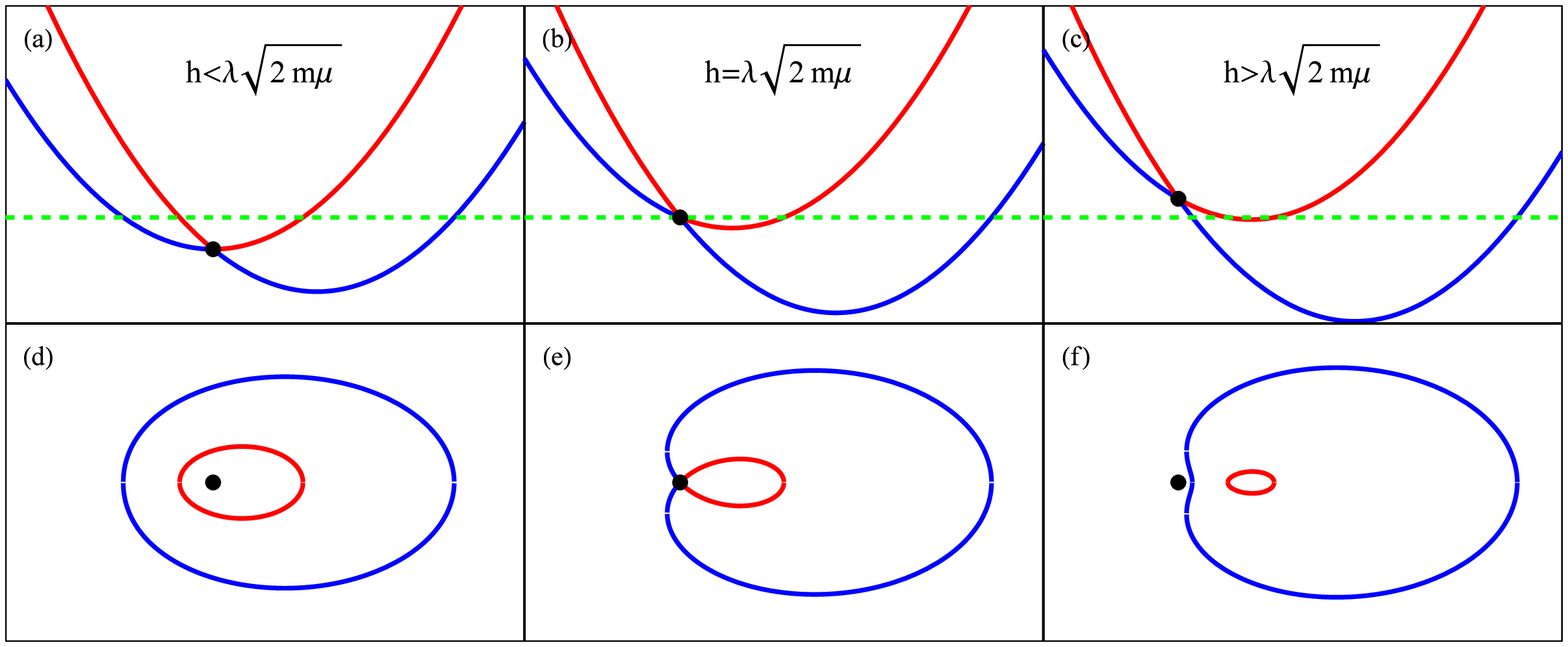}
\caption{(Color online) Upper panels: The dispersion relations $\ve_c(\bk_\perp=0,k_z)$. Lower panels: The Fermi-surface topologies in $(k_z,k_x)$ plane. The green dashed line represents the chemical potential. Blue (red) lines are for right-(left-)handed fermions. When $h<\l\sqrt{2m\m}$, the Berry monopole (the black point) locates inside both the Fermi surfaces; At $h=\l\sqrt{2m\m}$ the two Fermi surfaces touch; when $h>\l\sqrt{2m\m}$, the Berry monopole moves out both the Fermi surfaces.} 
\label{illu_top}
\end{center}
\vspace{0.cm}
\end{figure}

If the right-hand side does not vanish, \eq{anomaly} represents a quantum anomaly for the current of chirality $c$ in the form analogous to the chiral anomaly in gauge field theory. This can be seen more clearly if we consider a Fermi gas at zero temperature with pure Weyl SOC, $\bL=\l\bk$, where $\l$ is the strength of the SOC. In this case, $\bO=\bk/(2k^3)$, $\nabla_\bk\cdot\bO=2\p\d^{(3)}(\bk)$, $F=1$, $\bk_0={\bf0}$, and $f_c(\bk_0)=1$. Thus the right-hand side of \eq{anomaly} reads $c(\be\cdot\bb)/(4\p^2)$, which coincides exactly with the $U(1)$ chiral anomaly. In this case, the conservation of particle numbers of chirality $c$ which is proportional to the volume of its corresponding Fermi sphere is violated by the flux of the Berry curvature across the Fermi surface. For a Weyl-Zeeman SOC, $\bL(\bk)=\l \bk+h\hat{\bf z}$ with $h>0$ being a constant Zeeman field, the Berry curvature is
\begin{eqnarray}
\label{oweyl}
\bO&=&\l^2\frac{\bL(\bk)}{2W^3},
\end{eqnarray}
the Berry monopole locates at $\bk_0=-h/\l\hat{\bf z}$, and the winding number $F=1$. At zero temperature, we have
$f_c(\bk_0)=\h[\m-h^2/(2m\l^2)]$ which vanishes if the Zeeman field $h>\l \sqrt{2m\m}$. Thus the system has quantum anomaly only when $h<\l\sqrt{2m\m}$. The absence of the quantum anomaly when $h>\l \sqrt{2m\m}$ reflects a change of the Fermi-surface topology as shown in \fig{illu_top}.

We note here that there is no quantum anomaly for Rashba-Dresselhaus SOC $\bL(\bk)=(\l k_y-\w k_x, -\l k_x+\w k_y, h)$ with $\l, \w, h>0$ being constants, because its Berry curvature,
\begin{eqnarray}
\bO&=&\frac{(\l^2-\w^2)h}{2[h^2+(\l^2+\w^2)(k_x^2+k_y^2)-2\l\w k_x k_y]^{3/2}}\hat{\bf z},
\end{eqnarray}
leads to zero winding number.

Before we proceed, let us comment on the validity regime of the semiclassical approach. To arrive at \eqs{eomx6}{eomp6} the inter-band transition has been neglected which means that the force acting on the atoms cannot be strong. In particular, this require that $\sqrt{|\bb|}\ll W(\bp)/\l$; see Refs.~\cite{Sundaram,Xiao:2010,Stephanov} for more discussions. Obviously, this condition is violated if $\bk$ is close to the Berry monopole where $W(\bp)$ is small. Thus the phase space integral in \eq{anomaly} should be understood to exclude the region $|\bk-\bk_0|<\D$ around the Berry monopole with $\D$ large enough so that we can apply the classical description to particles outside of it. The value of $\D$ depends on $\bb$ and $\be$. For example, for pure Weyl SOC, $\D$ should be larger than $\sqrt{|\bb|}$; this actually constrains the magnitude of $\bb$: because $\D$ should not exceed $k_F$, the maximum $|\bb|$ should not exceed $k_F^2$. This implies that the rotation frequency should not exceed $\ve_F$ in order to guarantee the validity of the semiclassical approach. Our numerical simulations will always be within the validity regime of the semiclassical approach.

\begin{figure}
\begin{center}
\includegraphics[height=3.3cm]{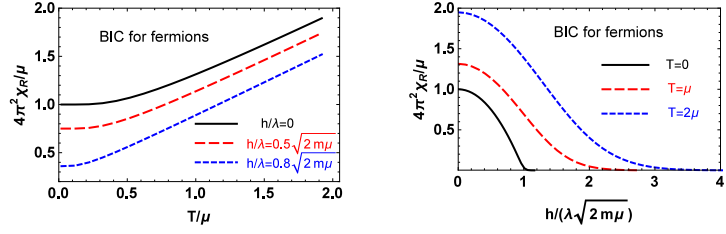}
\caption{(Color online) The $\bb$-induced conductivity, $\c_R$, for Fermi gases with Weyl-Zeeman SOC as function of the temperature (left panel) and $h/(\l \sqrt{2m\m})$ (right panel). Note that $\c_L=-\c_R$.}
\label{chi_fermi}
\end{center}
\vspace{0.cm}
\end{figure}

{\bf Chiral magnetic and chiral separation effects.---}
A consequence of this quantum anomaly is the appearance of CME and CSE. To see this,
we substitute \eq{eomx6} into \eq{current1},
\begin{eqnarray}
\label{current2}
\bj_c&=&\int\frac{d^3\bk}{(2\p)^3}f_c\nabla_\bk\ve_c+c\be\times\int\frac{d^3\bk}{(2\p)^3}\bO f_c\non
&&+c\bb\int\frac{d^3\bk}{(2\p)^3}(\bO\cdot\nabla_\bk\ve_c)f_c.
\end{eqnarray}
The first term on the right-hand side is the normal number current, the second term is the (intrinsic) anomalous Hall effect
and the last term represents a $\bb$-induced current which we denote by $\bj_c^{\rm \bb-ind}$:
\begin{eqnarray}
\label{current3}
\bj^{\rm \bb-ind}_c&=&\c_{c}\bb,\\
\label{chiexpr}
\c_{c}&=&c\int\frac{d^3\bk}{(2\p)^3}(\bO\cdot\nabla_\bk\ve_c)f_c,
\end{eqnarray}
where $\c_c$ is the $\bb$-induced conductivity (BIC) of chirality $c$. Let us consider $f_c$ to be the Fermi-Dirac distribution,
\begin{eqnarray}
\label{fd}
f_c= f_c^0=\frac{1}{e^{\b\ve_c}+1}\;\;\; {\rm for\; fermions}.
\end{eqnarray}
In this case, one finds that
\begin{eqnarray}
\c_R=-\c_L=\frac{T}{4\p^2}\ln\ls1+\exp{(-\b\ve_0)}\rs
\end{eqnarray}
with $\ve_0=\m[h^2/(2m\m\l^2)-1]$. Note that the $\l$ and $h$ dependence of $\c_c$ is through their ratio $h/\l$, and thus when $h=0$ the BIC is independent of $\l$ as long as it is nonzero. We present the numerical results for $\c_R$ in \fig{chi_fermi}.
It is seen that the BIC is enhanced at finite temperature and suppressed by large Zeeman field or small SOC strength. The latter effect is more transparent at zero temperature at which $f_c^0=\h(-\ve_c)$ and $\c_c$ can be analytically obtained:
\begin{eqnarray}
\label{chic}
\c_{c}(T=0)&=&\frac{c}{4\p^2}\lb\m-\frac{h^2}{2m\l^2}\rb\h\lb\mu-\frac{h^2}{2m\l^2}\rb.
\end{eqnarray}
Thus, the $\bb$-induced currents disappear when $h>\l\sqrt{2m\m}$ which again reflects a Fermi-surface topology transition (see \fig{illu_top}), in parallel to the absence of the quantum anomaly.

The above result $\c_R=-\c_L$ reflects the fact that the right-handed and left-handed atoms have the same chemical potential. If the atomic cloud contains domains (a possible realization will be presented in next section) in which the right-handed and left-handed chemical potentials differ, say,
\begin{eqnarray}
\m_R=\m+\m_A,\;\; \m_L=\m-\m_A,
\end{eqnarray}
the two BICs, $\c_R$ and $\c_L$, in these domains will also differ in magnitude and \eq{chic} becomes
\begin{eqnarray}
\label{chic2}
\c_{c}(T=0)
&=&\frac{c}{4\p^2}\lb\m_c-\frac{h^2}{2m\l^2}\rb\h\lb\mu_c-\frac{h^2}{2m\l^2}\rb.
\end{eqnarray}
At zero Zeeman field, by inserting $\c_{c}(T=0)$ into \eq{current3} we obtain the $\bb$-induced vector and chiral currents:
\begin{eqnarray}
\label{eqcme}
\bj_V^{\rm\bb-ind}&\equiv&\bj_R^{\rm\bb-ind}+\bj_L^{\rm\bb-ind}=\frac{\m_A}{2\p^2}\bb,\\
\label{eqcse}
\bj_A^{\rm\bb-ind}&\equiv&\bj_R^{\rm\bb-ind}-\bj_L^{\rm\bb-ind}=\frac{\m}{2\p^2}\bb.
\end{eqnarray}
These equations express the CME and the CSE in forms consistent with \eq{qcdcme}.

Finally, we note that the BICs do not receive perturbative corrections from scattering~\cite{Son2,Hou,Banerjee,Jensen,Satow} (See the Method.). This originates from the fact that the chiral anomaly is free of renormalization (the Adler-Bardeen theorem).

\begin{figure}
\begin{center}
\includegraphics[width=8.cm]{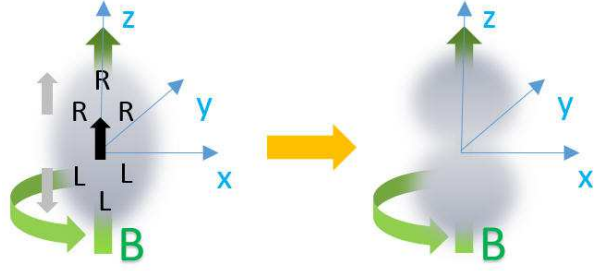}
\caption{(Color online) A schematic illustration of CME and CSE induced chiral dipole and mass quadrupole in Fermi gases with Weyl SOC. First, the CSE drives a chirality separation along the rotation axis with $\m_A>0$ in the upper tip and $\m_A<0$ in the lower tip (the left panel). Then the CME in turn drives particle number or equivalently the mass to flow away from the center and the atomic cloud acquires a mass quadrupole along the rotation axis (the right panel).}
\label{illu}
\end{center}
\vspace{0.cm}
\end{figure}

{\bf Chiral dipole and mass quadrupole.---}
Now we turn to the phenomenology of the CME and CSE. We consider a Weyl-Zeeman spin-orbit coupled Fermi gas in normal phase with $\m>0$. Once it is rotating, the CSE (\ref{eqcse}) will drive the chirality to move along $\bb$ and causes a macroscopic separation of the chirality, see \fig{illu} (left). This chiral dipolar distribution naturally forms two separated domains, one with $\m_A>0$ in the upper space while another with $\m_A<0$ in the lower space. Then in these domains the CME (\ref{eqcme}) in turn drives particle number or equivalently the mass to flow away from the center and the atomic cloud acquires a mass quadrupole along $\bb$, see \fig{illu} (right). In the context of heavy-ion collisions, similar mechanism was proposed to generate a charge quadrupole in QGP which may induce a difference between the elliptic flows of $\p^+$ and $\p^-$~\cite{shov,cmvexp} that has been detected at RHIC~\cite{Wangcmv}.

\begin{figure}
\begin{center}
\includegraphics[width=8.5cm]{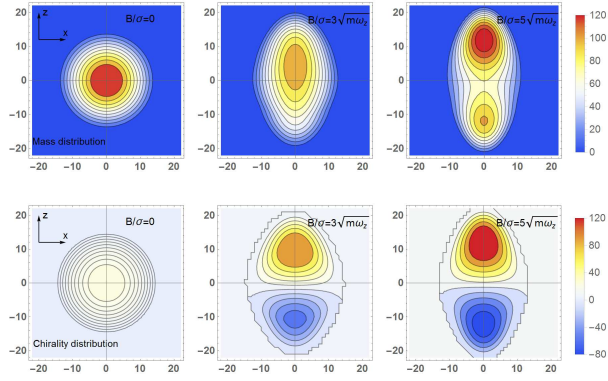}
\caption{(Color online) Mass quadrupole (upper panels) and chiral dipole (lower panels) induced by CME and CSE at different rotating frequencies. Length is in unit of $1/\sqrt{m\o_z}$.}
\label{separation}
\end{center}
\vspace{0.cm}
\end{figure}

The above argument is true only at the qualitative level, to reveal the real dynamical process, one needs to solve the coupled evolution problem of the right- and left-handed currents. Let us consider a pure Weyl SOC and the $T=0$ case. The general forms of the right-handed and left-handed currents should contain diffusion and normal conducting terms, so they read (the anomalous Hall terms vanish for pure Weyl SOC because of the time-reversal symmetry)
\begin{eqnarray}
\bj_R&=&\frac{\m_R}{4\p^2}\bb-D_R\nabla n_R +\s\be,\\
\bj_L&=&-\frac{\m_L}{4\p^2}\bb-D_L\nabla n_L +\s\be,
\end{eqnarray}
where $D_{c}$ is the diffusion constant and $\s$ is the normal conductivity, they are linked by the Einstein relations $\s=D_R(\pt n_R/\pt\m_R)=D_L(\pt n_L/\pt\m_L)$. Consider a small fluctuation $\d n_c$ in density $n_c$. This will induce a small departure of the chemical potential from the background value,
$\m_c=\m_{\rm bg}+\d\m_c$, where $\m_{\rm bg}=-A_0$ is linked to $\be$ by $\be=\nabla\m_{\rm bg}$. Substituting $\bj_{c}$ to \eq{anomaly} and keeping linear order terms in $\d n_c$ we obtain
\begin{eqnarray}
\frac{\pt\d n_R}{\pt t}+\frac{1}{4\p^2}\frac{\pt\m_R}{\pt n_R}\bb\cdot\nabla\d n_R-D_R\nabla^2\d n_R=0,\\
\frac{\pt\d n_L}{\pt t}-\frac{1}{4\p^2}\frac{\pt\m_L}{\pt n_L}\bb\cdot\nabla\d n_L-D_L\nabla^2\d n_L=0.
\end{eqnarray}
These two equations represent two wave modes with dispersions $E_R(\bk)=v_R k-iD_Rk^2$, $E_L(\bk)=v_L k-iD_Lk^2$,
one propagating along $\bb$ with velocity $v_R=B({\pt\m_R}/{\pt n_R})/(4\p^2)$ and another
 opposite to $\bb$ with velocity  $v_L=B({\pt\m_L}/{\pt n_L})/(4\p^2)$. We call them chiral magnetic waves (CMWs) in accordance with the same wave modes found in the context of QGP~\cite{cmv}. It is the CMWs that develops the chiral dipole and the mass quadrupole. Unlike the situation in QGP, in our setup the trapping potential will finally balance the driving force due to the CMWs and establish a new mechanical equilibrium. The new equilibrium will be characterized by the position-dependent chemical potential
\begin{eqnarray}
\m_c(\bx)&=&\mu-\tilde{V}(\bx)+\D\m_c(\bx),
\end{eqnarray}
where $\tilde{V}=(m/2)\lb\tilde{\o}_\perp^2(x^2+y^2)+\o_z^2 z^2\rb$ is the effective trapping potential and $\tilde{\o}_\perp^2=\o_\perp^2-\o^2$. The chemical potential shift $\D\m_c(\bx)$ is determined by the mechanical equilibration condition $\bj_R=\bj_L=0$. The numerical results are shown in \fig{separation} where we simulate $5\times10^5$
fermions in the harmonic trap. Length is in unit of $1/\sqrt{m\o_z}$, and we assume the transverse effective trapping frequency $\tilde{\o}_\perp$ being kept to be $\o_z$ when the $\bb$-field changes. Experimentally, the chiral dipole may be hard to detect but the mass quadrupole profile can be easily detected by, e.g., light absorption images.

{\bf Conclusion.}---
In summary, we have demonstrated that if the atomic gases with Weyl-Zeeman SOC is 1) trapped by an external potential and 2) under rotation, there can appear a quantum anomaly in the chiral currents. A consequence of this chiral anomaly is the chiral magnetic effect and chiral separation effect. The CME and
CSE cause macroscopic separation of chirality and a mass quadrupole along the rotation axis in the fermionic atomic cloud which may possibly be detected in cold atomic experiments.

In the context of QGP, there has been found other transport phenomena that stem from the quantum anomaly, e.g., the chiral electric separation effect~\cite{Huang2,Jiang}, which may also be realized in spin-orbit coupled atomic gases in the similar setup as we discussed in this Report. In addition, it is clear from the derivation that the chiral anomaly may exist also in Bose gases. How the quantum-anomaly-induced transports in bosonic gases also deserve detailed exploration in future works.

{\bf Method.}---
Now we show the robustness of $\bb$-induced currents against scattering. Let us consider the collision kernel $I[f_c]$ in the relaxation time approximation,
\begin{eqnarray}
I[f_c]=-\frac{\d f_c}{\t},
\end{eqnarray}
where $\d f_c=f_c-f_c^0\propto \bb$ (which is assumed to be small in this calculation) with $f_c^0$ the equilibrium distribution. We now show that $\c_c$ is independent of $\t$. The relaxation time $\t$ is assumed to be independent of the chirality, and we use it to characterize the interaction strength among atoms. Concretely, the relaxation time can be expressed as $\t=1/(n\bar{v}\s_T)$ with $n$ the density, $\bar{v}$ the average velocity, $\s_T=4\p a^2/(1+\bar{k}^2a^2)$ the total cross section, and $a$ the scattering length. For fermions, at low temperature, $\t\sim3\p m(1+(k_Fa)^2)/(4k_F^2(k_Fa)^2)$. At high temperature, $\t\sim m^{3/2}\sqrt{T}/n$.

To proceed, let us turn the $\be$-field off and assume a steady state with no spatial dependence of $f_c$ and $\bb$. Substituting $f_c$ into the kinetic equation (18) in the main text, at linear order in $\bb$, we obtain
\begin{eqnarray}
\d f_c\approx-\frac{\t}{\sqrt{G_c}}(\nabla_\bk\ve_c\times\bb)\cdot\nabla_\bk f_c^0.
\end{eqnarray}
The $\bb$-field induced current of chirality $c$ is given by
\begin{eqnarray}
\bj_c^{\bb-{\rm ind}}=\int\frac{d^3\bk}{(2\p)^3}\sqrt{G_c}\dot{\bx} (f_c^0+\d f_c),
\end{eqnarray}
where the first term gives the result (26) in the main text and the second term
\begin{eqnarray}
&&\int\frac{d^3\bk}{(2\p)^3}\sqrt{G_c}\dot{\bx} \d f_c\non&=&-\t\int\frac{d^3\bk}{(2\p)^3}\dot{\bx} (\nabla_\bk\ve_c\times\bb)\cdot\nabla_\bk\ve_c \frac{\pt f_c^0}{\pt \ve_c}=0.
\end{eqnarray}
Thus the $\bb$-induced currents are solely given by CME and CSE and are free of perturbative corrections.

{\bf Acknowledgments}
We acknowledge the support from Fudan University Grant No. EZH1512519, Shanghai Natural Science Foundation No. 14ZR1403000, the Key Laboratory of Quark and Lepton Physics (MOE) of CCNU (Grant No. QLPL20122), the Young 1000 Talents Program of China, and Scientific Research Foundation of State Education Ministry for Returned Scholars.

{\bf Author Contributions}\\
X.G.H conceived and conducted the research and wrote the manuscript. Correspondence and requests for materials
should be addressed to X.G.H. (huangxuguang@fudan.edu.cn).

{\bf Competing Interests}\\
The author declares that he has no competing financial interests.

\end{document}